\documentclass[10pt]{article}
\usepackage{graphicx}
\usepackage{amsmath,amssymb}
\usepackage{color}      
\usepackage{hyperref}
  \newcommand{\core}{{\rm core} }
   \newcommand{\MEPSatNLO}{{\sc\small  MEPS@NLO}}
      \newcommand{\MEPSatNLOQCD}{{\sc\small  MEPS@NLO~QCD}}
   \newcommand{\FxFx}{{\sc\small  FxFx} }

\newcommand\pubnumber{Cavendish-HEP-19/01,  IPPP/19/3,\\ NIKHEF/2019-001, TUM-HEP-1182/19,\\ TTK-19-01}

\def\institute{ \small $^{1}$Institut f\"ur Theoretische Teilchenphysik und Kosmologie,  RWTH Aachen University, D-52056 Aachen, Germany\\
  $^{2}$Department of Physics and Astronomy, University College London, Gower Street, London, WC1E 6BT, UK\\
  $^{3}$Institute for Particle Physics Phenomenology, Department of Physics, Durham University, Durham, DH1 3LE, UK \\
$^{4}$Cavendish Laboratory, University of Cambridge, Cambridge CB3 0HE, UK\\
$^{5}$Technische Universit\"{a}t M\"{u}nchen, James-Franck-Str.~1, D-85748 Garching, Germany\\
$^{6}$Theoretical Physics Department, CERN, CH-1211 Geneva 23, Switzerland\\
$^{7}$Nikhef, Science Park 105, NL-1098 XG Amsterdam, The Netherlands\\
}

\def\Title#1{\begin{center} {\Large #1 } \end{center}}
\def\Author#1{\begin{center}{ \sc #1} \end{center}}
\def\Address#1{\begin{center}{ \it #1} \end{center}}

\newcommand\pubblock{\rightline{\begin{tabular}{l} \pubnumber\\
           \end{tabular}}}
\newenvironment{Abstract}{\begin{quotation}  }{\end{quotation}}
\newenvironment{Presented}{\begin{quotation} \begin{center} 
             PRESENTED AT\end{center}\bigskip 
      \begin{center}\begin{large}}{\end{large}\end{center} \end{quotation}}
\def\Acknowledgements{\bigskip  \bigskip \begin{center} \begin{large}
             \bf ACKNOWLEDGEMENTS \end{large}\end{center}}

\def\beq{\begin{equation}}
\def\eeq#1{\label{#1}\end{equation}}
\def\eeqn{\end{equation}}

\def\beqa{\begin{eqnarray}}
\def\eeqa#1{\label{#1}\end{eqnarray}}
\def\eeqan{\end{eqnarray}}

\let\bar=\overbar

\def\Dslash{\not{\hbox{\kern-4pt $D$}}}
\def\dslash{\not{\hbox{\kern-2pt $\del$}}}

\def\msb{{\bar{\ssstyle M \kern -1pt S}}}

\begin{document}
\begin{titlepage}
\pubblock

\vfill
\Title{NNLO versus NLO multi-jet merging for top-pair production including electroweak corrections}
\vfill
\Author{ Micha\l{}  Czakon$^{1}$, Christian G\"utschow$^{2}$, Jonas M.~Lindert$^{3}$, \\ Alexander Mitov$^{4}$, Davide Pagani$^{5}$,  Andrew S.~Papanastasiou$^{4}$, \\ Marek Sch\"onherr$^{3,6}$, Ioannis Tsinikos$^{5}$, Marco Zaro$^{7}$}
\Address{\institute}
\vfill
\begin{Abstract}
In this proceedings we compare phenomenological predictions for differential distributions in top-quark pair production at the LHC. In particular we consider NNLO QCD fixed-order predictions and parton-level predictions based on NLO QCD multi-jet merging following the  MEPS@NLO scheme. In both predictions NLO electroweak (EW) corrections are incorporated in different approximations.
We focus on several  transverse-momentum distributions and on the top-quark invariant mass distribution, both
highly relevant for the ongoing physics program at the LHC.
We discuss comparisons between the  different considered approximations and their advantages and disadvantages for different distributions.
\end{Abstract}
\vfill
\begin{Presented}
$11^\mathrm{th}$ International Workshop on Top Quark Physics\\
Bad Neuenahr, Germany, September 16--21, 2018\\
by Davide Pagani
\end{Presented}
\vfill
\end{titlepage}
\def\thefootnote{\fnsymbol{footnote}}
\setcounter{footnote}{0}

\section{Introduction}
In this proceedings we compare phenomenological predictions for differential distributions in top-quark pair production at the LHC with 13 TeV, using two different approximations: fixed-order NNLO QCD combined with electroweak (EW) corrections at NLO, and  multi-jet merging at NLO also incorporating EW corrections. The latter is applicable for predictions at the particle level.

The NNLO QCD predictions for top-pair production are based on the results of Refs.~\cite{Czakon:2013goa,Czakon:2015owf}, which have been combined with  EW corrections in Refs.~\cite{Czakon:2017wor,Czakon:2017lgo}. Besides the well known NLO EW corrections of relative 
$\mathcal{O}(\alpha\alpha_S^2)$~\cite{Kuhn:2006vh,Bernreuther:2006vg,Kuhn:2013zoa,Bernreuther:2010ny,
Hollik:2011ps,Pagani:2016caq} these combinations further included all subleading Born and one-loop contributions, 
first considered in~\cite{Pagani:2016caq} and based on the {\sc\small MadGraph5\_aMC@NLO}  framework \cite{Alwall:2014hca, Frederix:2018nkq}.

The NLO multi-jet merged predictions are based on the {\MEPSatNLO} framework in {\sc\small Sherpa}~\cite{Hoeche:2009rj,Hoeche:2012yf,Gehrmann:2012yg} and
EW corrections are incorporated in an approximation integrating out QED radiation effects at the level of the NLO
calculation~\cite{Kallweit:2015dum}. Predictions for top-pair production merged up to one extra jet at NLO QCD+EW
and 2 additional jets at LO in this approximation have been presented in~\cite{Gutschow:2018tuk}, which serves as basis for the {\MEPSatNLO} 
predictions considered in this proceedings. These predictions will be denoted as {\MEPSatNLO}, while multi-jet merged predictions without EW corrections will be denoted as {\MEPSatNLOQCD}. In Ref.~\cite{Gutschow:2018tuk} also full fixed-order one-loop predictions have been presented for top-pair and top-pair production in association with an extra jet. It was found that the relative EW corrections behave universily between the two processes, indicating a factorization of QCD and EW higher-order corrections.

We focus on different transverse-momentum $(p_T)$ distributions and on the top-quark invariant mass. In particular, we consider the $p_T$ of the top quark and anti-quark, $p_T(t)$ and $p_T(\bar t)$, of the leading and trailing top, $p_T(t_1)$ and $p_T(t_2)$, and their average value at the histogram level, $p_{T,\rm{avt}}\equiv(p_{T}(t)+p_{T}(\bar t))/2=(p_{T}(t_1)+p_{T}(t_2))/2$.

\section{Input parameters and scales}
In order to exclude other possible sources of discrepancies in the comparison, we have used for both the approximations the same input parameters, which we conventionally set equal to those of Ref.~\cite{Czakon:2017wor}. Also the same PDF set {\sc\small NNPDF3.1luxQED}  \cite{Bertone:2017bme} has been used, which includes a photon density \`a la {\sc\small LUXqed} \cite{Manohar:2016nzj,Manohar:2017eqh}. Although the calculation of Ref.~\cite{Gutschow:2018tuk} does not include photon-induced contributions, it has been shown in Ref.~\cite{Czakon:2017wor} that with this set of PDFs they are at the permille level and therefore negligible for the comparison in this work.

For what concerns the choice of the central value of the factorisation and renormalisation scales, different choices, both physically motivated, are employed in the two different calculations. In the case of NNLO predictions their central values $\mu$ depend on the distribution that is considered,
\begin{eqnarray}
\mu &=& \frac{m_{T}(t)}{2}~~{\rm for~the} ~ p_T(t) ~ {\rm distribution}, \label{eq:scalemT}\\
\mu &=& \frac{m_{T}(\bar t)}{2}~~{\rm for~the} ~ p_T(\bar t) ~ {\rm distribution}, \label{eq:scalemTbar}\\
\mu &=& \frac{H_T}{4} = \frac{1}{4} \left( m_{T}(t)+ m_{T}(\bar t) \right)~~{\rm for~all~other~distributions,}
\label{eq:scaleHT}
\end{eqnarray}
and have been identified in Ref.~\cite{Czakon:2016dgf} via the ``Principle of Fastest Convergence''. Instead, in the multi-jet merging approximation, a CKKW scale  \cite{Catani:2001cc,Bothmann:2016nao} has been used, with a scale $\mu_\core$ for the $pp\rightarrow t \bar t$ hard process set equal to 
\begin{equation}\label{eq:corescale}
    \mu_\core
    =\frac{1}{2} \left(\frac{1}{\hat s}+\frac{1}{m_t^2-\hat t}+\frac{1}{m_t^2-\hat u} \right)^{-\frac{1}{2}}\, ,
\end{equation}
which in practice weights the contributions of the different colour flows involved in the process \cite{Hoeche:2013mua,Hoeche:2014qda}.

It is very interesting to notice that
\begin{align}
{\rm for}~p_T(t)\rightarrow\infty  \qquad   &\mu_{\rm core}\Longrightarrow \frac{1}{2}\sqrt{\frac{4}{5}}p_{T}\sim\frac{p_{T}}{2}\,, \label{eq:boost}\\
{\rm for}~E(t)\rightarrow \infty {~\rm and}~p_T(t)/E(t)\rightarrow 0 \qquad &\mu_{\rm core}\Longrightarrow \frac{m_{T}}{2}\sim\frac{H_{T}}{4}\,, \label{eq:highmass}\\
{\rm for}~E(t) \rightarrow 0 \qquad &\mu_{\rm core}\Longrightarrow \frac{1}{2}\sqrt{\frac{4}{5}}m_{t}\sim\frac{H_{T}}{4}\,, \label{eq:threshold}
\end{align}
where $E(t)$ is the energy of the top quark. In other words, even though the scale choice in the two different calculations have been identified on the basis of very different principles, they are very close in the full phase-space: in the boosted regime \eqref{eq:boost}, at high $m(t\bar t)$ \eqref{eq:highmass},\footnote{At high $m(t \bar t)$, top (anti)quarks are mainly produced in the peripheral region, due to $t$- and $u$-channel diagrams, therefore $p_T(t)/E(t)\rightarrow 0$.} and at the threshold \eqref{eq:threshold}. Morever, \eqref{eq:boost}-\eqref{eq:threshold} suggest that the observable-dependent scale choices in \eqref{eq:scalemT}-\eqref{eq:scaleHT} can actually be derived from a single definition at the fully differential level.
\section{Results}
\begin{figure}[t]
\centering
\includegraphics[width=0.48\textwidth]{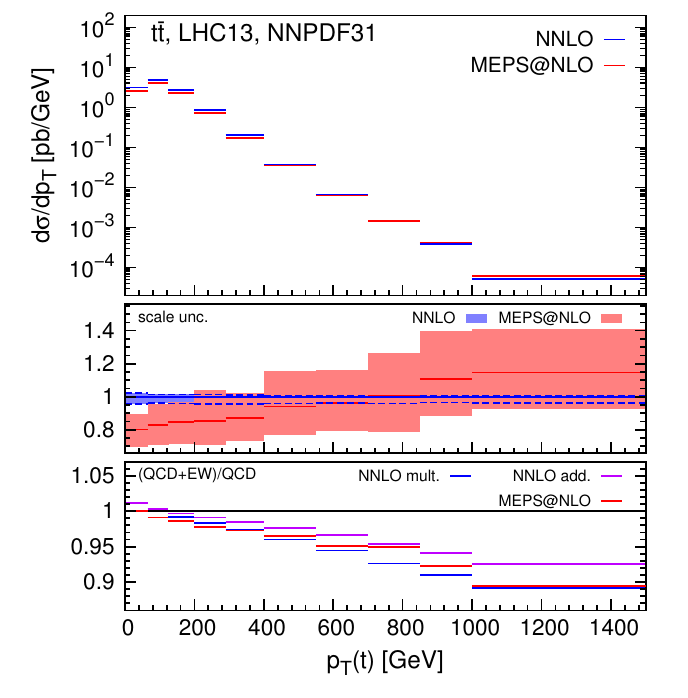}
\includegraphics[width=0.48\textwidth]{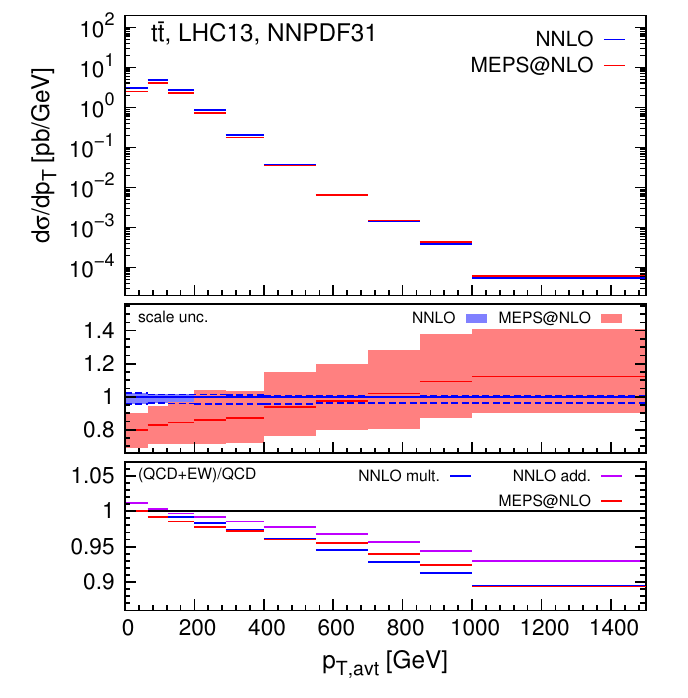}
\caption{Comparison between {\MEPSatNLO} and NNLO predictions for the $p_{T}(t)$ and $p_{T,\rm{avt}}$.}
\label{fig:ptta}
\end{figure}
\begin{figure}[t]
\centering
\includegraphics[width=0.48\textwidth]{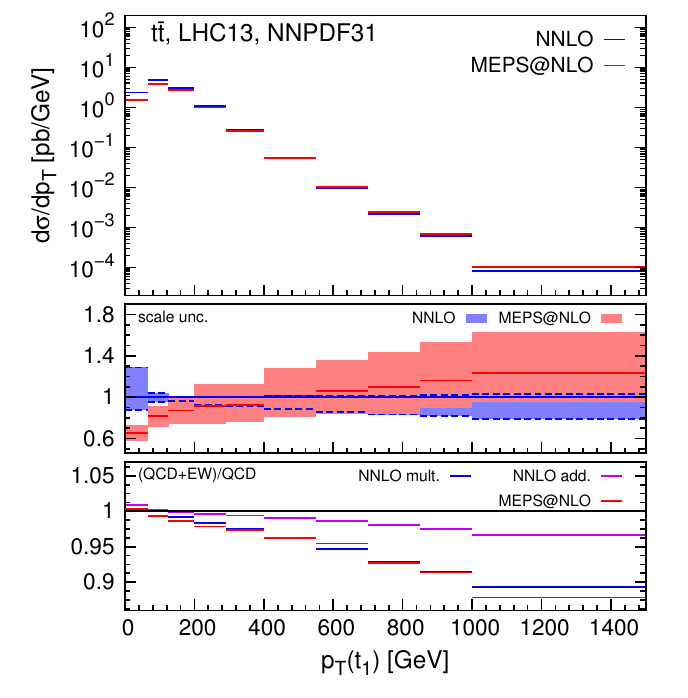}
\includegraphics[width=0.48\textwidth]{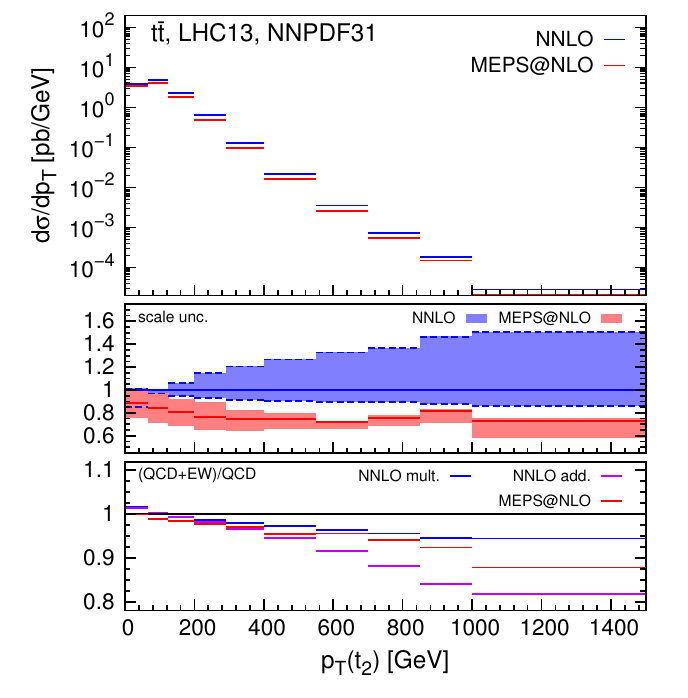}
\caption{ Comparison between {\MEPSatNLO} and NNLO predictions for the $p_{T}(t_1)$ and $p_{T}(t_2)$. 
}
\label{fig:pt12}
\end{figure}
In Fig.~\ref{fig:ptta} we compare NNLO and {\MEPSatNLO} predictions for $p_{T}(t)$ and $p_{T,\rm{avt}}$ distributions. In the first inset we show the scale uncertainties~\footnote{At NNLO we have used 7-point variation, while at {\MEPSatNLO} the 9-point variation. However, the off-diagonal values are within the uncertainty band and thus this difference has no impact.} for the two different approximations, both normalised over the central value of the NNLO one. As can be seen, besides the first bin, for both distributions the uncertainty bands from the two different predictions overlap. As expected, the scale uncertainty at NNLO is much smaller than in the {\MEPSatNLO} predictions. The information about the EW corrections is instead displayed in the second inset.  We show, for both NNLO and {\MEPSatNLO}, the ratio of their predictions with and without including EW contributions,  at their central scales. Moreover, for the NNLO, we consider also the case in which we have combined QCD and EW approach in the additive approach, see Ref.~\cite{Czakon:2017wor} for details. These plots further support the multiplicative approach, since the {\MEPSatNLO} value is much closer to the NNLO in the multiplicative approach than in the additive one.

Using the same layout, in Fig.~\ref{fig:pt12} we compare NNLO and {\MEPSatNLO} predictions for $p_{T}(t_1)$ and $p_{T}(t_2)$ distributions. As can be seen, there are large discrepancies between the two approximations. The reason is that these two distributions are pathological at fixed order. Indeed, since $p_{T}(t_1)>p_{T}(t_2)$, with fixed order calculations  a jet-veto on additional radiation is indirectly applied at small values of $p_{T}(t_1)$ and leads to large uncontrolled terms for $p_{T}(t_1) \lesssim m_t$. On the contrary, adding shower effects and thus multiple radiations, this effect is automatically cured in {\MEPSatNLO}. The same argument applies for $p_{T}(t_2) \gtrsim m_t$. Comparing the impact of the electroweak corrections is therefore not meaningful in these phase-space regions. Outside these two regions ($p_{T}(t_1) \gtrsim m_t$ or $p_{T}(t_2) \lesssim m_t$), where fixed-order calculations are reliable, the two calculations are again compatible and NNLO scale uncertainty is much smaller than at {\MEPSatNLO}. Moreover, as can be seen in Fig.~\ref{fig:pt12}, the impact of electroweak corrections in {\MEPSatNLO} and NNLO (in the multiplicative approach) is again very similar. We remark that averaging the distributions in the plots on the left and the right in Fig.~\ref{fig:pt12} the $p_{T,\rm{avt}}$ distributions of Fig.~\ref{fig:ptta} is obtained, where all these issues are not present.
\begin{figure}[t]
\centering
\includegraphics[width=0.47\textwidth]{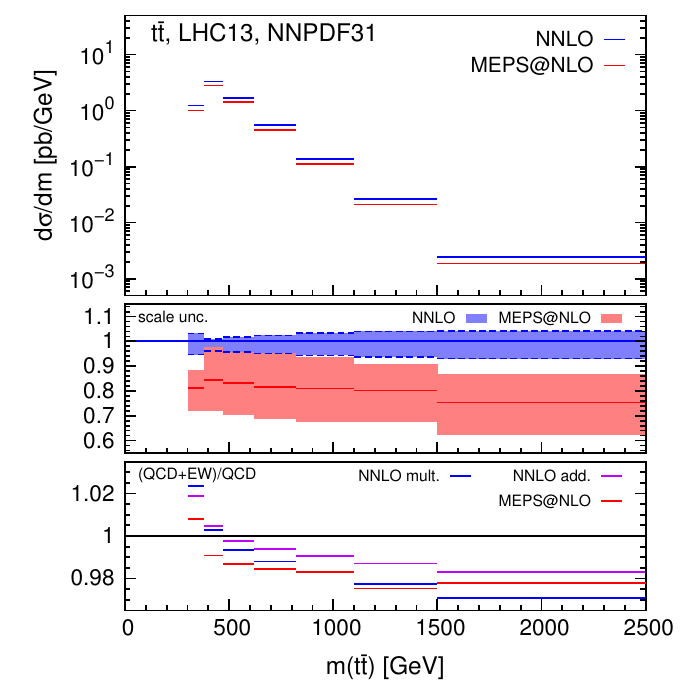}
\includegraphics[width=0.5\textwidth]{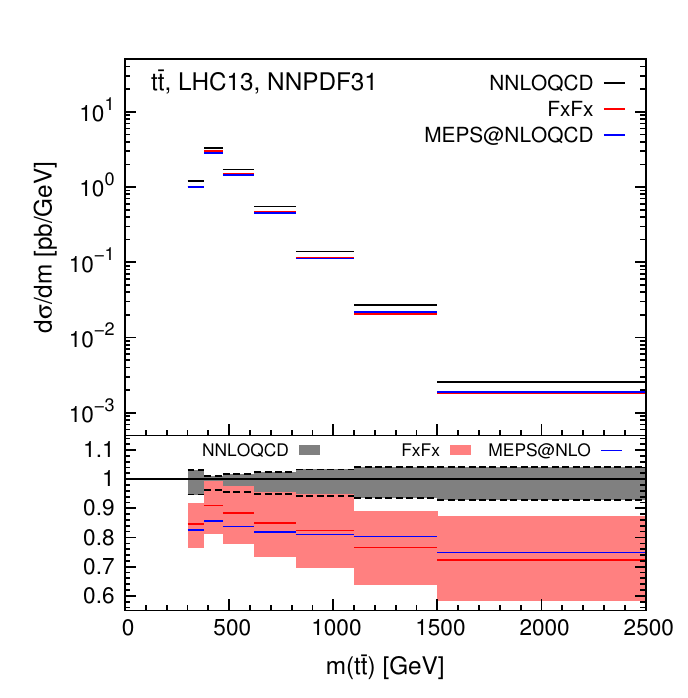}
\caption{ Comparison between {\MEPSatNLO} and NNLO predictions (left) and between purely QCD {\MEPSatNLO}, \FxFx and NNLO predictions (right) for the $m(t \bar t)$ distribution. 
}
\label{fig:mtt}
\end{figure}

In the left plot in Fig.~\ref{fig:mtt} we show the same kind of plot for the $m(t \bar t)$ distribution. As can be seen, in this case the agreement is far to be perfect, especially in the tail of the distribution. We have checked that this is not due to the particular procedure that is used in the NLO merging in {\MEPSatNLO} by comparing the purely QCD predictions with the one obtained with \FxFx \cite{Frederix:2012ps}, which also provides NLO-merged predictions with a different procedure. The plot on the right of Fig.~\ref{fig:mtt} shows this comparison. We have also verified that setting the scales exactly to the same values in the NNLO and {\MEPSatNLO} approaches reduces but does not completely eliminate this tension. Further work is therefore necessary in order to fully understand the origin of this discrepancy between the two approaches. Both approaches correctly take into account the contributions from one hard real emission at one-loop  and two hard real emissions at tree level; a possible origin of the discrepancy may be due to the missing two-loop terms in the {\MEPSatNLO} and/or the missing shower effects in the NNLO.

\section{Conclusions and Outlook}
In this proceeding we have compared two different approximations for phenomenological predictions of top-quark pair production at the LHC: NNLO QCD combined with EW corrections at NLO, and {\MEPSatNLO} multi-jet merging at NLO also including EW corrections. We have considered different distributions at 13 TeV:  $p_{T}(t)$, $p_{T,\rm{avt}}$, $p_{T}(t_1)$, $p_{T}(t_2)$ and $m(t \bar t)$. In the case of $p_{T}(t)$ and  $p_{T,\rm{avt}}$, the two approximations are compatible and, as expected, NNLO predictions have much smaller scale uncertainties. Thus, this approximation is more suitable for precision studies on parton-level distributions. On the other hand, fixed-order calculations can be pathological and indeed this is the case for $p_{T}(t_1)$ and $p_{T}(t_2)$ distributions in the regions $p_{T}(t_1) \lesssim m_t$ and $p_{T}(t_2) \gtrsim m_t$, respectively, where the {\MEPSatNLO} is superior to fixed-order calculations. 

We have also shown that the relative effects induced by EW corrections in {\MEPSatNLO} is much closer to those observed at NNLO with EW corrections combined in the multiplicative approach than in the case when they are simply added (additive approach).

For the specific case of $m(t \bar t)$ a tension between the two approaches is present, especially in the far tail, therefore further work is necessary in order to fully understand its origin. Additional plots regarding this aspect can be found at the repository:\\

\href{http://www.precision.hep.phy.cam.ac.uk/results/ttbar-nnloqcd-nloew/}{\texttt{http://www.precision.hep.phy.cam.ac.uk/results/ttbar-nnloqcd-nloew/}}
\\

We conclude that an NNLO+PS calculation would be desirable for precise predictions in the full phase-space and especially for future studies. For this purpose, we remark the relevance of EW corrections and we have further supported the superiority of the multiplicative approach for their combination with those from QCD origin, especially in the boosted regime.

\Acknowledgements

The work of M.C.~has been support in part by BMBF. The work of A.M.~and A.P.~is supported by the European Research Council Consolidator Grant NNLOforLHC2 and by the UK STFC grants ST/L002760/1 and ST/K004883/1. The work of D.P.~and I.T.~is supported by the Alexander von Humboldt Foundation, in the framework of the Sofja Kovalevskaja Award Project ``Event Simulation for the Large Hadron Collider at High Precision''. M.S.~acknowledges the support of the Royal Society through the award of a University Research Fellowship. The work of MZ is supported by the Netherlands National Organisation for Scientific Research (NWO).

\end{document}